\RequirePackage{fix-cm}
\documentclass[smallextended]{svjour3}       

\smartqed  
\usepackage{graphicx}
\usepackage{tikz}
\usepackage{amsmath}
\usepackage{amssymb}
\usepackage{commath}
\usepackage{xcolor}
\usepackage{listings}

\begin{document}

\title{Landau: language for dynamical systems with automatic
differentiation}


\author{Ivan Dolgakov         \and
        Dmitry Pavlov 
}


\institute{Institute of Applied Astronomy of the Russian Academy of Sciences \\
          Russia, 191187, St. Petersburg, Kutuzova Embankment, 10 \\
          Tel.: +7 (812) 275-1118\\
          Fax: +7 (812) 275-1119\\
          Corresponding e-mail: \texttt{ia.dolgakov@iaaras.ru}
}

\date{Received: date / Accepted: date}

\maketitle

\begin{abstract}
  Most numerical solvers used to determine free variables of dynamical systems rely on first-order
  derivatives of the state of the system w.r.t. the free variables. The number of the free variables can be
  fairly large. One of the approaches of obtaining those derivatives is the integration of the 
  derivatives simultaneously with the dynamical equations, which is best done with the automatic 
  differentiation technique.\\
  Even though there exist many automatic differentiation tools, none have been found to be scalable
  and usable for practical purposes of dynamic systems modeling. Landau is a Turing incomplete 
  statically typed domain-specific language aimed to fill this gap. The Turing incompleteness provides 
  the ability of sophisticated source code analysis and, as a result, a highly optimized compiled code. 
  Among other things, the language syntax supports functions, compile-time ranged for loops, if/else 
  branching constructions, real variables and arrays, and the ability to manually discard calculation 
  where the automatic derivatives values are expected to be negligibly small. In spite of reasonable 
  restrictions, the language is rich enough to express and differentiate any cumbersome paper-equation 
  with practically no effort.
\end{abstract}

\section{Introduction}
\label{intro}
In dynamical system modeling, various systems from different application
domains can be represented by an autonomous system of first-order ODEs:
\begin{equation}\label{eq:ODE}
  \dot{\vec{x}}(t) = f(\vec x(t), \vec p).
\end{equation}
where $\vec p = \{p_i\}_{i=1}^m$ is a vector of $m$ fixed parameters.
One instance of the model is based on the values of the parameters, and also
the initial conditions:
\begin{equation}\label{eq:initial}
  \vec{x}(t_0) = \vec{x}_0
\end{equation}

For example, in case of an N-body dynamical system the parameters are masses, and the initial 
conditions are positions and velocities at a certain moment of time.
In practice, e.g. in planetary ephemerides, the precise values of the initial conditions are unknown, 
while some approximate values are determined from observations.

The task is to solve the initial value problem (IVP) (\ref{eq:ODE}, \ref{eq:initial}) for a range of $t$ 
covering all the $\{t_i\}$ and to minimize the discrepancy between observed
and computed values. The IVP is most often solved numerically.

Let $\vec{P} = (x_0^{(0)},...,x_0^{(n)}, p_1,...,p_m)$ be the full set of
free variables to be fit to observations by (as usually done with dynamical systems)
nonlinear least squares method. The first-order derivatives
$\frac{\mathrm{d}\vec x}{\mathrm{d}\vec P}$ are required for the method.

One way to obtain $\frac{\mathrm{d}\vec x}{\mathrm{d}\vec P}$ is to include
it into our system of ODEs, together with $\vec x$ itself. Accordingly,
the initial conditions $\frac{\mathrm{d}\vec x_0}{\mathrm{d}\vec P}$
and the time derivative  $\frac{\mathrm{d}}{\mathrm{d}t}\od{\vec{x}}{\vec{P}}$
are needed to solve the IVP for the new system.
While the initial conditions are trivial, the time derivative must be
obtained by substituting~(\ref{eq:ODE}):
\begin{equation}
  \frac{\mathrm{d}}{\mathrm{d} t}\od{\vec{x}}{\vec{P}} = \od{f(\vec{x}, \vec p)}{\vec{P}}.
\end{equation}

Thus, in order to estimate the free variables, one needs to compute the derivative of the ODE's right-hand side w.r.t. $\vec{P}$. 
There are three ways to perform such a computation:

\begin{itemize}
  \item Full symbolic differentiation, which requires a computer algebra system and can
  be quite computationally costly.
  \item Numeric differentiation using the the finite difference technique, which is prone to truncation errors.
  \item Automatic differentiation.
\end{itemize}

Automatic Differentiation (AD) is a technique of obtaining numerical values of derivatives of a given 
$\mathbb{R}^n \rightarrow \mathbb{R}^m$ function (listing~\ref{code:func}). As opposite to the 
symbolic differentiation, AD not only reduces the computation time by using memoization 
techniques but also provides more flexibility as it can deal with complicated structures from 
programming languages, such as conditions and loops. Because of the chain rule associativity there are at 
least two ways (modes) of memoization: forward and reverse. 

The forward mode of AD is based on the concept of dual numbers and on traversing the computational graph 
(fig. \ref{fig:func_comp_graph}) in natural forward order. Each variable of the original program is associated with its 
derivative counterpart(s), which is(are) computed along with the original variable value (see listing~\ref{code:forward}). 
The computational complexity of forward mode is proportional to the number of independent input variables $n$, thus it 
is most effective when $n \ll m$. In our practice the number of the function output values is far greater then 
the number of the input ones, therefore we used forward accumulation.

\begin{figure}[!htb]
  \centering
  \begin{minipage}{.45\textwidth}
  \begin{lstlisting}[caption={A function with $n = 2$, $m = 1$: the Kepler equation $M = E - e \sin(E)$}, captionpos=b, label=code:func]
func $M$($E$, $e$):
  $w_1$ = $E$
  $w_2$ = $e$
  $w_3$ = sin((*$w_1$*))
  $w_4$ = $w_3 * w_2$
  $w_5$ = $w_1 - w_4$
  $M$ = $w_5$
  return($M$)
      \end{lstlisting}
  \end{minipage}%
  \hspace{0.05\textwidth}
  \begin{minipage}{0.45\textwidth}
      \centering
      \begin{tikzpicture}[scale=.8, label distance=-4pt]
        \begin{scope}[every node/.style={circle,thick,draw}]
            \node[label=left:$w_1$] (x1) at (0,0) {$E$};
            \node[label=left:$w_2$] (x2) at (4,0) {$e$};
            \node[label=left:$w_3$] (sin) at (1.8,1.2) {$\sin$};
            \node[label=left:$w_4$] (*) at (3,2) {$*$};
            \node[label=left:$w_5$] (-) at (2,4) {$-$};
            \node (f) at (2,6) {$M$};
            
        \end{scope}
    
        \begin{scope}[>=stealth,
                      every node/.style={fill=white,circle},
                      every edge/.style={draw=black,thick}]
            \path (x1) edge (-);
            \path (x1) edge (sin);
            \path (x2) edge (*);
            \path (sin) edge (*);
            \path (*) edge (-);
            \path (-) edge (f);
            
        \end{scope}
    \end{tikzpicture}
      \caption{Computational graph of function from listing~\ref{code:func}.}
      \label{fig:func_comp_graph}
  \end{minipage}
\end{figure}
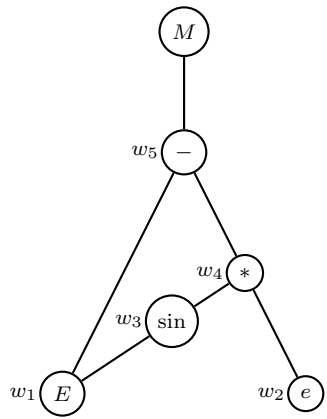

\begin{figure}
  \begin{minipage}{.45\textwidth}
    \begin{lstlisting}[caption={Taking the derivatives $\pd{M}{E}$ and $\pd{M}{e}$, using the forward AD mode}, captionpos=b, label = {code:forward}]
func d$M$($E$, $e$):
  $\pd{w_1}{E}$ = 1
  $\pd{w_1}{e}$ = 0
  $w_1$ = $E$
  
  $\pd{w_2}{E}$ = 0
  $\pd{w_2}{e}$ = 1
  $w_2$ = $e$
  
  $\pd{w_3}{E}$ = cos($w_1$) * $\pd{w_1}{E}$
  $\pd{w_3}{e}$ = cos($w_1$) * $\pd{w_1}{e}$
  $w_3$ = sin($w_1$)

  $\pd{w_4}{E}$ = $\pd{w_3}{E}$ * $w_2$ + $w_3$ * $\pd{w_2}{E}$
  $\pd{w_4}{e}$ = $\pd{w_3}{e}$ * $w_2$ + $w_3$ * $\pd{w_2}{e}$
  $w_4$ = $w_3$ * $w_2$
  
  $\pd{w_5}{E}$ = $\pd{w_1}{E}$ - $\pd{w_4}{E}$
  $\pd{w_5}{e}$ = $\pd{w_1}{e}$ - $\pd{w_4}{e}$
  $w_5$ = $w_1$ - $w_4$
  
  $\pd{f_1}{E}$ = $\pd{w_5}{E}$
  $\pd{f_2}{E}$ = $\pd{w_5}{e}$
  $M$ = $w_5$
  return($M$, $\pd{M}{E}$, $\pd{M}{e}$)
    \end{lstlisting}
  \end{minipage}
  \hspace{0.05\textwidth}
  \begin{minipage}{.45\textwidth}
    \begin{lstlisting}[label = {code:reverse}, caption={Taking the derivatives $\pd{M}{E}$ and $\pd{M}{e}$, using the reverse AD mode}]
func d$M$($E$, $e$):
  $w_1$ = $E$
  $w_2$ = $e$
  $w_3$ = sin((*$w_1$*))
  $w_4$ = $w_3 * w_2$
  $w_5$ = $w_1 - w_4$
  $M$ = $w_5$

  $\pd{M}{w_5}$ = 1
  $\pd{M}{w_4}$ = $\pd{M}{w_5}$ * (-1)
  $\pd{M}{w_3}$ = $\pd{M}{w_4}$ * $w_2$
  $\pd{M}{w_2}$ = $\pd{M}{w_4}$ * $w_3$
  $\pd{M}{w_1}$ = $\pd{M}{w_5}$ * 1 + $\pd{f}{w_3}$ * cos($w_1$)
  $\pd{M}{e}$ = $\pd{M}{w_2}$ * 1
  $\pd{M}{E}$ = $\pd{M}{w_1}$ * 1
  return($M$, $\pd{M}{E}$, $\pd{f}{e}$)
            \end{lstlisting}
  \end{minipage}
\end{figure}
   
In the case of reverse mode values of the derivatives are accumulated from the root(s) of the computational graph, 
each assignation is augmented with $m$ accumulations (see listing~\ref{code:reverse}). 
Hence, the computation complexity of reverse mode is proportional to the number of the function outputs $m$; 
thus, it is most effective when $n \gg m$, which is often the case in computation of gradients
of many-to-one function so widely used in the neural networks.

\section{Related work and motivation}
There exist a large number of forward-mode AD software tools for
differentiating functions that are written in general-purpose programming
languages, like Fortran (ADIFOR)~\cite{bischof1992adifor}, C
(ADIC)~\cite{bischof1997adic} or C++ (ADOL-C)~\cite{griewank1996algorithm}.
Rich features of the ``host'' languages, like arrays, loops, conditions, and
recursion, often make it difficult to implement a practically usable AD system
without imposing limitations on the language and/or extra technical work when
specifying the function, especially in presence of multi-dimensional
functions with many independent variables.

On the other hand, there exist a number of languages developed specially
for AD tasks, like Taylor~\cite{JorbaZou} and VLAD~\cite{Siskind2008,Siskind2016}.
Taylor syntax, while very simple and natural, is very limited (no conditionals,
loops, arrays, or subprocedures). VLAD, a functional Scheme-like language, has
conditionals, loops, recursion, and subprocedures, but does not have arrays
or mutability.

Finally, there are tools for differentiating functions specified as mathematical
expressions in mathematical computing systems, like MATLAB (ADMAT)~\cite{coleman1998admat}
or Mathematica (TIDES)~\cite{TIDES,TIDES2}.
Such tools often require a bigger effort (as compared to a general-purpose
languages) to input a practical dynamical system of large dimension with a lot
of free variables.

In this work, a new language, Landau, is proposed, designed specially for
dynamical systems. Other examples of such design are TIDES and Taylor.
However, TIDES and Taylor obtain high-precision solutions using Taylor method
and high-order derivatives, while Landau provides only first-order derivatives
and is supposed to be used with numerical integrators that obtain an acceptable
approximate solution (like Runge-Kutta or Adams methods), with better
performance than high-precision methods.

Like VLAD, Landau is a domain-specific language designed with automatic
differentiation in mind. Like TIDES and Taylor, Landau offers C code
generation. Like general-purpose languages, Landau has common control flow
constructs, arrays, and mutability; but unlike general-purpose languages, Landau
embraces Turing incompleteness to perform static source analysis (see section
\ref{sec:impl}) and generate efficient code.

Landau also has the ability to not only derive derivative dependencies from
source (e.g. if $x = y + z$, then $\pd{x}{y} = 1 + \pd{z}{y}$), but also to fix
values of derivatives to other variables belonging to the dynamical system
(e.g. $\pd{y}{a} = \text{var}$).

\section{Syntax}
The language syntax offers functions, mutable real and integer variables, mutable real arrays, 
constants, \texttt{if/else} statements and \texttt{for} loops. Special type \texttt{parameter} is used 
to express Jacobian denominator variables which are not used in expressions (the right-hand sides of the 
assignments) itself. In case of dynamical equations differentiation such parameters could express 
initial conditions vectors. Special derivative operator \texttt{'} is used to annotate or assign the 
value of the derivative. Even with branching constructions (\texttt{if/else} statements) the function 
is guaranteed to be continuously differentiable thanks to the prohibition of the real arguments inside the 
condition body. Moreover, it is allowed to manually omit negligibly small derivatives 
using the \texttt{discard} keyword (e.g. if $x(a) = y(a) + z(a) + t(a)$ and command \texttt{discard y ' a} 
is typed, then $\pd{x}{a} = \pd{x}{a} + \pd{t}{a}$).

Listing \ref{code:spacecraft} demonstrates a Landau program for a
dynamical system describing the motion of a spacecraft.
The state of the system, i.e. the 3-dimensional position and velocity of the
spacecraft, obeys Newtonian laws. The derivatives of the state
w.r.t. 6 initial conditions (position and velocity) and one parameter
(the gravitational parameter of the central body) are calculated using AD.

\begin{lstlisting}[caption={Landau program for modeling spacecraft movement around 
  a planet. Spacecraft's initial position and velocity, as well as the gravitational parameter of the 
  planet, are supposed to be determined by nonlinear least-squares method}, captionpos=b, escapeinside={
  (*}{*)}, label = {code:spacecraft}]
#lang landau 

# $\text{Annotated parameters. Function does not have them directly}$
# $\text{as arguments, but has derivatives w.r.t. them in the state vector.}$
parameter[6] initial

real[6 + 36 + 6] x_dot (
  real[6 + 36 + 6] x, # $\text{state + derivatives w.r.t. initial and GM}$
  real GM)
{
  real[36] state_derivatives_initial = x[6 : 6 + 36]
  real[6] state_derivatives_gm = x[6 + 36 : ]
  real[6] state = x[ : 6]

  # $\text{Set the state vector's Jacobian values.}$
  state[ : ] ' initial[ : ] = state_derivatives_initial[ : ]
  state[ : ] ' GM = state_derivatives_gm

  real[6] state_dot
  # $\text{Transfer the time derivatives from x to their xdot counterparts,}$
  # $\text{because }\dot x = v_x.$
  state_dot[ : 3] = state[3 : ]

  # $\text{Write the}$ state_dot $\text{part to the function output.}$
  x_dot[ : 3] = state_dot[ : 3]
 
  # $\text{Apply Newtonian laws.}$
  real dist2 = sqr(state[0]) + sqr(state[1]) + sqr(state[2])
  real dist3inv = 1 / (dist2 * sqrt(dist2))

  state_dot[3 : ] = GM * (-state[ : 3]) * dist3inv

  # $\text{Write the}$ state_dot $\text{part to the function output.}$
  x_dot[3 : ] = state_dot[3 : ]
  
  # $\text{Write the}$ state_dot $\text{derivatives to the function output.}$
  x_dot[6 : 6 + 36] = state_dot[ : ] ' initial[ : ]
  x_dot[6 + 36 : 6 + 36 + 6] = state_dot[ : ] ' GM
}
\end{lstlisting}

\section{Implementation}\label{sec:impl}
Automatic differentiation can be implemented in one of two ways: the operator overloading and the 
source code transformation. The first approach is based on describing the dual number data structure 
and overloading arithmetic operators and functions to operate on them. 
The second one involves analysis of function source and generation of the differentiation code. It was 
found~\cite{tadjouddine2002ad} that the latter approach generally produces more efficient derivative 
code. Landau is written in Racket~\cite{racket} and it uses source code transformation approach to produce Racket or ANSI 
C differentiation code.

Let \texttt{lvalue} be the variable in the left-hand side of assignation and \texttt{rvalues} be the variables in the 
right side. The differentiation is performed in the following way: each real \texttt{lvalue} is associated with an 
array carrying derivatives' values. The right part of the assignment is differentiated 
symbolically\footnote{Even though the reverse mode is truly preferable if $n > m$, which is the 
case in term level assignment, because there is only one output in each assignation (e.g. $m = 1$), 
the computation overhead is negligibly small in case of small expressions.}, the result is carried in 
the accumulator array.

Each real variable assignation in a forward scheme is augmented with $n$ assignations to the derivatives' 
accumulators, but in practice there is no need to compute and store all of them because some 
derivatives' values are never used afterwards. That means that the computed Jacobians are often sparse.

To illustrate the sparsity problem and keep things simple let us consider an artificial migration problem 
over $N$ areas with a simplified diffusion model of migration:
\begin{eqnarray*}
  \frac{\mathrm{d}p_i}{\mathrm{d}t}& =& \sum_{\substack{j = 0\\ j \neq i}}^{N} m_{ij} p_j, \quad i \in [0, N),  \\
   p_j(t_0) & = & p_j^{(0)}, \quad j \in [0, N)
\end{eqnarray*}
where the initial condition vector $\mathbf{p}^{(0)} = \{p_j\}^{(0)}$ is supposed to
be determined from observational data.
Say that there are $k$ regions with $l = \frac{N}{k}$ strongly interconnected areas whose population at an arbitrary
moment of time depends from the initial conditions of other areas within that region. Following the logic from 
the introduction, we need to find the solution derivatives with respect to the initial conditions. 
The Landau program for solving that problem is presented in listing~\ref{code:migration}. 

\begin{lstlisting}[caption={Example Landau program demonstrating the sparsity. One-to-many migration in k = 10 regions over the N = 1000 areas.}, label = {code:migration}]
#lang landau
const int N = 1000
const int k = 10
const int l = N / k
const int L2 = l * l
parameter[N] p0

real[N + L2 * k] f
  (real[N * N] m,
   real[N]     p,
   real[N * N] derivatives_p0) {

  p[ : ] ' p0[ : ] = derivatives_p0[ : ]

  real[N] p_dot
  for i = [0 : N]
    for j = [0 : N]
      if (i != j) {
        p_dot[i] += m[N * i + j] * p[j]
      }
  
  f[0 : N] = p_dot[ : ]

  for i = [0 : k]
    f[N + L2 * i : N + L2 * i + L2] = 
      p_dot[l * i : l * i + l] ' p0[l * i : l * i + l]
}
  \end{lstlisting}

The fact that the population depends on the initial conditions only within the 
region makes the Jacobian $\od{\mathbf{p}\phantom{_0}}{\mathbf{p}_0}$ sparse:

\begin{equation}
  \begin{bmatrix}
    J_{0,0}  & \cdots & J_{0,99} & & &  &  \\
    \vdots   & \ddots & \vdots & & & 0 &   \\
    J_{99,0} & \cdots & J_{99,99}& \\
    & & &  \ddots & & \\
    & & & &  J_{900,900} & \cdots & J_{900,999} \\
    & 0 & & &  \vdots       & \ddots & \vdots \\
    & & & & J_{999,900}  & \cdots & J_{999,999} \\
  \end{bmatrix}.
\end{equation}
In the following simple example the sparsity pattern is presented with square blocks on the main diagonal but
it could be randomly sparse in general. Accumulating the $p$'s derivatives in a straightforward manner will require one
to compute and store $N^2$ values while only $\frac{N^2}{k}$ are needed. 

There are at least two approaches to deal with sparsity. The first approach is to generate the code where each 
useful\footnote{We are not using term nonzero here, because it can happen that the useful Jacobian matrix element is equal to zero.} 
 Jacobian matrix element is stored in a separate variable. That involves unrolling loops 
to the assignation sequences and, as a result, facing the performance penalty due to the CPU cache 
misses. Another approach is to store useful Jacobian values in arrays and preserve the ability to use 
loops for traversing. The sparsity is handled by packing useful Jacobian elements to smaller arrays and 
generating mappings from the packed derivative indexes to the original ones and inverse mappings, which map the original indexes to the packed ones. The listing~\ref{code:diff_pseudo} demonstrates the 
differentiation of the loops from the lines 16--20 of the listing~\ref{code:migration}.

The compilation is performed in two stages. During the first stage the
information about dependencies, used variables and derivatives is gathered for
each variable or array cell. The Turing incompleteness guarantees that all loops
and conditions can be unrolled and computed at compile time, thus the initial
Landau function can be transformed to a list of actions (listing~\ref{code:actions_list}): 
derivative annotation, variable assignation and storage of the derivative in the output value. 
The list is then traversed to gather the dependency graph of the derivatives, which is used in the second 
compilation stage to generate mappings, inverse mappings and differentiation code.

\begin{figure}[!htb]
  \begin{lstlisting}[caption={Pseudocode of the loops (lines 16--20 of the listing~\ref{code:migration}) differentiation}, label = {code:diff_pseudo}]
for $i$ in [$0$ : $N$]:
  for $j$ in [$0$ : $N$]:
    if $i \neq j$:
      for $k$ in mappings$_{\texttt{p\_dot, p0}}$(i):
        dp_dot_dp0[$k$] = dp_dot_dp0[$k$] 
          + m[$N$ * $i$ + $j$] * dp_dp0[inv_mapping$_{\texttt{p, p0}}$($j, k$)]
      p_dot[$i$] = p_dot[$i$] + m[$N$ * $i$ + $j$] * p[$j$]
  \end{lstlisting}
\end{figure}

\begin{figure}[!htb]
  \begin{lstlisting}[caption={Reversed actions list generated from the listing~\ref{code:migration}}, label = {code:actions_list}]
need-this-derivative p_dot[999] ' p0[999]
need-this-derivative p_dot[999] ' p0[998]
need-this-derivative p_dot[999] ' p0[997]
need-this-derivative ...
...
p_dot[999] depends-from {p_dot[999], p[998]}
p_dot[999] depends-from {p_dot[999], p[997]}
p_dot[999] depends-from {p_dot[999], p[996]}
...        depends-from ...
...
have-this-derivative p[0] ' p0[2]
have-this-derivative p[0] ' p0[1]
have-this-derivative p[0] ' p0[0]
    \end{lstlisting}
\end{figure}

Let $H$ be the length of parameter vector and $h_x$ be the number of derivatives (e.g. Jacobian row's 
elements) needed for the $x$ variable. When $h_x \ll H$, most Jacobian values are not used and thus 
should not be computed. Using the mappings technique described above we store only $h_x$ derivative 
values and use mappings $l \rightarrow k$ and inverse mappings $k \rightarrow l$, where 
$l \in [0, h_x), k \in [0, H)$ to set and get derivative values. Mappings can be easily 
 implemented as arrays with length $h_x$ by storing the original indexes of the parameter vector. But 
 it is challenging to implement effective inverse mappings, because storing them in array directly will 
 result to the $\Theta(H_{\text{max}})$ memory consumption, where $H_{\text{max}}$ is the maximum used 
 parameter vector index. For example, even if one needs to compute derivative with respect to the last 
 parameter index, the resulting mapping is array of size 1, but inverse mapping's length is $H$.

More sophisticated way to implement inverse mappings is to use minimal perfect hash functions (MPHF). A 
perfect hash function maps a static set of $h$ keys into a set of $g$ integer numbers without 
collisions, where $g \ge h$. If $g = h$, the function is called minimal. Various asymptotically 
effective algorithms for generating such functions exist~\cite{botelho2007simple}, but it is not clear 
if the constant factors are small enough to make the generation of many MPHFs during the single compilation 
practically effective. In the current version of Landau the inverse mappings are implemented as integer arrays 
and thus are not quite memory-efficient.

\section{Conclusion}
A new language called Landau has been invented to fill the niche of a
domain-specific language designed for practically usable
forward-mode AD for estimating the values of free parameters of
a complex dynamical system.

A compiler that translates Landau code into either Racket or high-performance C
code, has been implemented, making the overall procedure of estimating free
variables fast and fluent.

Further work is required for more effective implementation of the inverse
mappings. Such an implementation clearly should be possible thanks to
Turing-incompleteness of Landau code that allows for complete static analysis.

\section{Acknowledgements}
Authors are thankful to their colleague Dan Aksim for reading drafts of this
paper and to Matthew Flatt and Matthias Felleisen of the PLT Racket team for
their help with the Racket programming platform.


\end{document}